\documentclass{aa}  

\usepackage{graphicx}
\usepackage{txfonts}
\begin{document}

   \title{Dynamical Constraints on a Population of Massive Interstellar Objects}
   \author{Oem Trivedil
          \inst{1}
          \and
          Abraham Loeb\inst{2}\fnmsep
          }

   \institute{Department of Physics and Astronomy, Vanderbilt University, Nashville, TN 37235, USA\\
              \email{oem.trivedi@vanderbilt.edu}
         \and
             Astronomy Department, Harvard University, 60 Garden St., Cambridge, MA 02138, USA  \\
             \email{aloeb@cfa.harvard.edu}
             }

   \date{\today}

 
  \abstract
   {The detection of kilometer-scale interstellar objects (ISOs) places strong dynamical constraints on their underlying population and distribution.}
   {We aim to assess whether the observed encounter rates of large ISOs can be explained by their expected velocity distribution, or whether anomalous velocity anisotropies are required.}
   {We first derive the cumulative encounter rate scaling with size from brightness-limited detection models. We then employ the Eddington inversion method to link power-law density profiles $\rho(r)\propto r^{-k}$ with phase-space distributions, illustrating how steep slopes imply strong velocity-space focusing. Finally, we develop a Liouville mapping formalism based on energy and angular momentum conservation, propagating the interstellar distribution inward to the detection region while explicitly incorporating gravitational focusing and anisotropy.}
   {We show that encounter rates of large ISOs require flux enhancements far beyond  expectations, that steep density profiles produce the necessary inward velocity bias, and that Liouville mapping provides a physically self-consistent way to reproduce the observed size dependent detection rates. The main results are framed in the context of the parameters for 3I/ATLAS, but the implications are general and go on to sharpen the distinction between natural dynamical mechanisms and potential artificial origins for ISOs.}
   {}

   \keywords{Interstellar Objects --
                3I/ATLAS --
                Non-Natural Trajectories
               }

   \maketitle
%

\section{Introduction}
The discovery of the first two interstellar objects (ISOs), 1I/' Oumuamua \cite{oumumeech2017brief} in 2017 and 2I/Borisov in 2019 \cite{borguzik2020initial} marked a turning point in planetary science, providing the first direct probes of small bodies formed outside the Solar System. Despite their common interstellar provenance these objects displayed strikingly divergent properties, with 1I/'Oumuamua being photometrically inactive yet exhibited statistically significant non-gravitational acceleration, while 2I/Borisov showing unmistakable cometary activity. These contrasting characteristics ignited a broad range of theoretical explanations.
\\
\\
The recent discovery of 3I/ATLAS \cite{seligman2025discovery}, with an orbital eccentricity $e \sim 6.1$, an inclination of $175^\circ$, and a heliocentric excess velocity of $V_\infty \sim 58 \,\text{km s}^{-1}$, adds a third case to this growing sample of ISOs. Early observations reveal that 3I/ATLAS hosts a faint but detectable coma, yet exhibits minimal rotational light curve variation and a moderately red spectral slope of $\sim 17\%$/100 nm. Its absolute magnitude near $H_V \approx 12.4$ suggests a nucleus radius in the range below 2.8 km \cite{2.8jewitt2025hubble} and up to 23 km \cite{23lisse2025spherex}, depending on the relative contribution of the coma and the nucleus to the brightness. Altogether, 3I/ATLAS represents the largest interstellar object observed to date.
\\
\\
The combination of its large inferred size, high incoming velocity, and specific trajectory presents significant dynamical challenges. Unlike smaller ISOs which may be ejected more readily from planetary systems during early dynamical instabilities, the presence of massive interstellar bodies at observable number densities like this strains standard models of ejection efficiency and population statistics. As with 1I/`Oumuamua and 2I/Borisov, we are hence led to consider the detection of 3I/ATLAS in a broader framework. Much work has hence followed to understand various properties of this object \cite{loeb20253i,a2hibberd2025interstellar,a3loeb2025intercepting,3iatde2025assessing,hopkins2025different}, as was the case for 1I/'Ouamuamua \cite{oum1loeb2022possibility,oum2forbes2019turning,oum3siraj2022mass,oum4siraj20192019,oum5bialy2018could}.One of the central questions is whether the appearance of a large object within the inner solar system is consistent with expected dynamical pathways, or does it point to the need for a revised understanding of either the ISO size distribution or their delivery mechanisms?
\\
\\
In this work, we build upon the population level arguments articulated in the rarity of 3I/ATLAS observation in \cite{loeb20253i} and quantify the dynamical constraints implied by the detection of 3I/ATLAS. Specifically, we evaluate the phase-space conditions and angular momentum distributions that would allow an object of its inferred size to be delivered through the observed trajectory, and we calculate the likelihood of such events under both natural and artificial population hypotheses. By considering both the small radius and large radius interpretations of 3I/ATLAS, we demonstrate how the rarity of its trajectory places stringent constraints on the abundance of massive interstellar objects. This analysis provides a new perspective on the emerging census of ISOs and directly addresses the question of whether the detection of 3I/ATLAS is dynamically compatible with a natural origin, or whether alternative explanations such as artificial or directional delivery need to be considered.

\section{Encounter rate analysis}
Since direct detections of ISOs remain rare, the statistical rate of encounters provides a powerful indirect diagnostic of the underlying ISO population. In particular, encounter rate formulations allow us to quantify the scaling of detection probabilities with object size, velocity distribution, and survey sensitivity. This is especially interesting in the case of an object like 3I/ATLAS, as it could highlight the enhancements in flux required to reconcile the observed ISO detections with natural population models.
\\
\\
We begin with the standard expression for the detection rate of ISOs, written as an encounter rate
\begin{equation}
\Gamma_1(R) = n(R) \cdot \langle v \rangle \cdot \sigma(R),
\end{equation}
where $n(R)$ is the number density of objects within a narrow bin $dR$ of radius $R$, $\langle v \rangle$ is the average encounter speed of interstellar objects relative to the Sun, and $\sigma(R)$ is the effective detection cross section for an object of radius $R$. Rather than modeling the cross section geometrically, we instead adopt a brightness-based detection model, following a similar formulation as in \cite{seligman2025discovery}. In this approach, the detection cross section is set by the maximum distance at which an object of size $R$ is detectable, given a fixed limiting apparent magnitude $m_{\text{lim}}$. Using standard photometric relations for reflected sunlight, the apparent magnitude of an object at heliocentric distance $r\gg 1~{\rm au}$, geocentric distance $\Delta$, and radius $R$, for a fixed geometric albedo $p$, is given approximately by
\begin{equation}
m \approx H + 5 \log_{10}(r \Delta),
\end{equation}
where the absolute magnitude $H$ itself depends on the object radius as
\begin{equation}
H = 17.1 - 2.5 \log_{10} \left( \frac{p}{0.04} \right) - 5 \log_{10}(R).
\end{equation}
Assuming constant albedo and phase behavior, we can express the apparent magnitude as
\begin{equation}
m \propto -5 \log_{10}(R) + 5 \log_{10}(r \Delta).
\end{equation}
Fixing the apparent magnitude to the detection threshold $m_{\text{lim}}$, we can then invert the above expression to solve for the maximum distance at which an object of size $R$ can be detected. This leads to
\begin{equation}
r_{\text{max}}(R) \propto R^{1/2},
\end{equation}
and hence, since the detection cross section scales with the observable area (a circle of radius $r_{\text{max}}$), we have
\begin{equation}
\sigma(R) \propto r_{\text{max}}^2 \propto R.
\end{equation}
To determine the number density $n(R)$, we assume that the ISOs follow a differential size distribution of the form
\begin{equation}
\frac{dN}{dR} \propto R^{-q}.
\end{equation}
Assuming spatial uniformity, the number of objects per unit volume in a size range $[R, R + dR]$ is
\begin{equation}
n(R) dR = \frac{dN}{dV},
\end{equation}
and so
\begin{equation}
n(R) dR \propto R^{-q} dR,
\end{equation}
which implies the differential number density scales as
\begin{equation}
n(R) \propto R^{-q}.
\end{equation}
Now, rather than compute the differential detection rate, we derive the cumulative detection rate which is the rate of encountering all ISOs with radius greater than or equal to $R$. To do so, we integrate the differential rate from $R$ to $\infty$. Since $\Gamma_1(R') \propto n(R') \cdot \sigma(R')$, and we have just shown that $n(R') \propto R'^{-q}$ and $\sigma(R') \propto R'$, the integrand becomes $R'^{-(q - 1)}$. The cumulative encounter rate is
\begin{equation}
\Gamma_1(R) \propto \int_R^{\infty} R'^{-(q - 1)} dR'.
\end{equation}
This evaluates (for $q > 2$) to
\begin{equation}
\Gamma_1(R) \propto R^{-(q - 2)}.
\end{equation}
Finally, including the average encounter speed $\langle v \rangle$, which is computed from the shifted Maxwellian distribution as
\begin{equation}
\langle v \rangle = \left( \frac{2}{\sqrt{\pi}} \right) \sigma_s \left( 1 + \frac{v_\odot^2}{6 \sigma_s^2} \right),
\end{equation}
with $\sigma = 25 \, \text{km/s}$ $v_\odot = 17.4 \, \text{km/s}$. We have taken the velocity dispersion $\sigma_s$ to be of this value, given recent results showing that the dispersion in the range $20-30$ km/s regime is reasonable. For example, \cite{kohandel2020velocity} showed a velocity dispersion in the region $23-28$km/s region, \cite{eubanks2021interstellar} discussed a value of 26.14 for local stellar population as well, while other recent works have also also considered this regime \cite{hopkins2025different,hopkins2025predicting}. The final expression for the cumulative detection rate becomes
\begin{equation}
\Gamma_1(R) = k \cdot R^{-(q - 2)} \cdot \langle v \rangle.
\end{equation}
To fix the normalization, we impose the condition for 3I/ATLAS,
\begin{equation}
\Gamma_1(0.6) = 0.2~\text{yr}^{-1}
\end{equation}
which gives
\begin{equation}
k = \frac{0.2}{(0.6)^{q - 2} \cdot \langle v \rangle}.
\end{equation}
Thus, the full expression becomes
\begin{equation}
\Gamma_1(R) = \frac{0.2}{\langle v \rangle} \cdot (0.6)^{q - 2} \cdot R^{-(q - 2)} \cdot \langle v \rangle = 0.2 \cdot \left( \frac{R}{0.6} \right)^{-(q - 2)}.
\end{equation}

This allows us to define the flux enhancement factor as
\begin{equation}
\frac{\Gamma_2}{\Gamma_1} \approx \left( \frac{R}{0.6} \right)^{q - 2} ,
\end{equation}
which gives the required enhancement in flux beyond the natural population in order to explain the detection rate of objects larger than a given radius $R$. This formulation is based on cumulative detection rate considerations and incorporates both the size distribution and the average velocity derived from the velocity distribution.
\\
\\
To determine the proportionality constant $k$, we use the result from \cite{loeb20253i} for 3I/ATLAS and require that the cumulative detection rate for all objects of radius greater than $R_0 = 0.6~\text{km}$ matches the observed detection rate of $\Gamma_1(0.6) = 0.2 \, \text{yr}^{-1}$.
\begin{figure}[!h]
    \centering
    \includegraphics[width=1\linewidth]{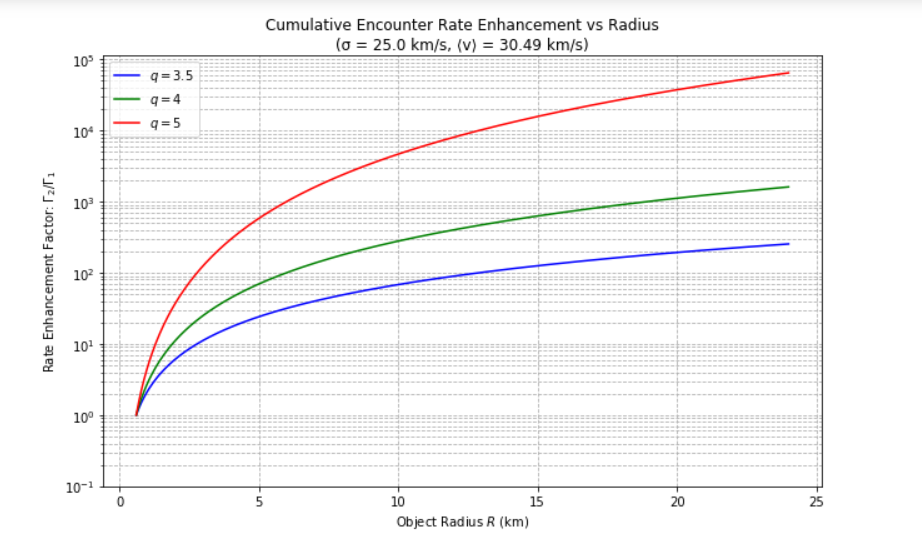}
    \caption{Cumulative encounter rate enhancement factor as a function of minimum ISO radius $R$ for different power-law slopes $q$, showing the steep increase in required enhancement for larger objects. The adopted parameters reflect representative values for 3I/ATLAS.}
    \label{25}
\end{figure}

Figure \ref{25} shows the cumulative encounter rate enhancement demonstrating how the expected detection frequency of interstellar objects depends on their radius and on the slope of the size distribution $q$. Note that we have considered a wide range of possible radii for 3I/ATLAS, from $\sim 2.8$km \cite{2.8jewitt2025hubble} to 23 km \cite{23lisse2025spherex}. While the natural component of the encounter rate sets the baseline expectation, the curves reveal that for larger radii the required enhancement factor rises steeply with increasing $q$ and in particular for steep size distributions ($q \gtrsim 4$), objects in the multi-kilometer regime demand encounter rates that exceed natural expectations by orders of magnitude. This might suggest an artificial or non-Maxwellian component in the detection statistics of large ISOs such as 3I/ATLAS \cite{seligman2025discovery}.

\section{Eddington-Inversion Analysis in Phase-Space}
The Eddington inversion method provides a powerful framework to connect spatial density profiles with the underlying velocity distribution functions in a self-consistent manner \cite{binney2011galactic,ed1lacroix2018anatomy,ed2suarez2010applicability}. For ISOs this approach is particularly useful as it naturally incorporates gravitational focusing by the Sun and reveals how an observed radial overdensity can emerge from specific velocity-space structures. By inverting an assumed power-law density profile, one directly obtains the phase-space distribution function $f(\mathcal{E})$, which determines the energy dependence of ISO orbits near the Sun.  
\\
\\
This not only allows us to identify the kinds of velocity biases required to reproduce enhanced detection rates, such as those observed for 3I/ATLAS, but also provides a means of distinguishing between shallow, weakly focused populations and steep, radially concentrated ones. We use the Eddington inversion method to determine the phase-space distribution function $f(\mathcal{E})$ that would yield a radial density profile $\rho(r) \propto r^{-k}$ under the Sun’s gravitational potential $\Psi(r) = GM_\odot / r$. This method self-consistently accounts for gravitational focusing and allows us to understand what kind of velocity distribution is needed to reproduce the enhanced detection rate of ISOs like 3I/ATLAS near the Sun.
\\
\\
We begin by assuming a power-law density profile for the incoming ISO population \cite{pl1napier2021capture,pl2siraj2019identifying,pl3penarrubia2023halo} in the gravitational potential of the Sun
 \begin{equation}
     \Psi(r) = \frac{GM_\odot}{r}
   \quad \text{so that} \quad
   \frac{d\Psi}{dr} = -\frac{GM_\odot}{r^2}.
 \end{equation}
We now use the Eddington inversion formula
\begin{equation}
    f(\mathcal{E}) = \frac{1}{\sqrt{8}\pi^2} \left[ \int_0^{\mathcal{E}} \frac{d^2\rho}{d\Psi^2} \frac{d\Psi}{\sqrt{\mathcal{E} - \Psi}} \right],
\end{equation}
where $\mathcal{E} = \Psi(r) - \frac{1}{2} v^2$ is the relative energy per unit mass. We now invert $\rho(\Psi) \propto \Psi^{k}$ by re-parametrizing our assumed $\rho(r)$ in terms of $\Psi$, then perform the differentiation and integration to compute $f(\mathcal{E})$.
\\
\\
We assume a power-law density profile
\begin{equation}
    \rho(r) \propto r^{-k}, \quad \text{with }~\Psi(r) = \frac{GM}{r} \Rightarrow r = \frac{GM}.{\Psi}
\end{equation}
Substitute into $\rho(r)$
\begin{equation}
    \rho(\Psi) \propto \left( \frac{GM}{\Psi} \right)^{-k} \propto \Psi^{k},
\end{equation}
and so
\begin{equation}
    \frac{d\rho}{d\Psi} \propto k \Psi^{k-1}, \quad
\frac{d^2\rho}{d\Psi^2} \propto k(k-1) \Psi^{k-2}.
\end{equation}
We this plug this into the Eddington formula to get 
\begin{equation}
    f(\mathcal{E}) \propto \int_0^{\mathcal{E}} \Psi^{k-2} \frac{d\Psi}{\sqrt{\mathcal{E} - \Psi}}
\end{equation}
Let $\Psi = \mathcal{E} u \Rightarrow d\Psi = \mathcal{E} du$, and $\mathcal{E} - \Psi = \mathcal{E}(1 - u)$ and then
\begin{equation}
    f(\mathcal{E}) \propto \mathcal{E}^{k - 3/2} \int_0^1 u^{k-2} (1 - u)^{-1/2} \, du
\end{equation}
This integral yields a Beta function solution as
\begin{equation}
    \int_0^1 u^{a-1} (1 - u)^{b-1} du = B(a, b) = \frac{\Gamma(a)\Gamma(b)}{\Gamma(a + b)}.
\end{equation}
The final result comes out to be (as the Beta function just acts as a constant)
\begin{equation}
    f(\mathcal{E}) \propto \mathcal{E}^{k - 3/2}.
\end{equation}
This yields a direct analytic expression for the phase-space distribution function as a function of the relative energy $\mathcal{E}$, parameterized by $k$. This $f(\mathcal{E})$ gives the required energy distribution for the ISO population that results in a radial density enhancement $\rho(r) \propto r^{-k}$ and since this formulation assumes spherical symmetry, the enhancement we get is entirely due to velocity-space bias towards the Sun.
\begin{figure}[!h]
    \centering
    \includegraphics[width=1.1\linewidth]{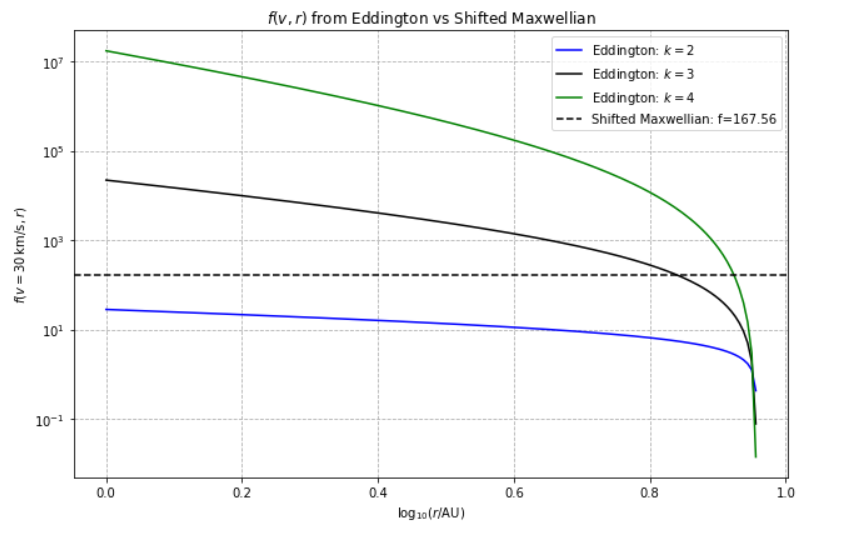}
    \caption{Comparison of Eddington-inverted phase-space densities for $k = 3, 3.5, 4$ with a shifted Maxwellian background.}
    \label{fvr}
\end{figure}
Figure \ref{fvr} shows the Eddington-inverted phase-space density $f(v = 30\,\mathrm{km/s}, r)$ as a function of $\log_{10}(r/\mathrm{AU})$ for different power-law density profiles alongside the value expected from a shifted Maxwellian background 
\begin{equation}
    f(v) \propto v^2 \exp\left( -\frac{v^2}{2\sigma^2} \right).
\end{equation}
Adopting for Sun's velocity relative  to the Local Standard of Rest $v = 30\,\text{km/s}$ and $\sigma = 25 \,\text{km/s}$ yields $f(v) \sim 167.56~({\rm km/s})^{-1}$. Note that this is just the background expected ISO flux at that velocity, assuming an isotropic, unperturbed distribution. Figure \ref{fvr} illustrates that the curves for $k = 3, 3.5, 4$ all exhibit a strong enhancement in $f(v, r)$ near the Sun (small $r$) and fall off rapidly at large distances. This behavior indicates a phase-space overdensity consistent with a population of ISOs concentrated on radial orbits toward the inner solar system. Such steep $k$-values are physically motivated when trying to explain the detection of large ISOs. By contrast, shallower profiles like $k = 1$ or $2$ imply only weak gravitational focusing and fail to generate the necessary enhancement in $f(v, r)$, underestimating the encounter rate at small heliocentric distances. The relation $f(\mathcal{E}) \propto \mathcal{E}^{k - 3/2}$ means that larger values of $k$ require a stronger concentration of low-angular-momentum orbits. Therefore, the steep Eddington profiles better reflect the degree of inward directional bias required to explain the observational data.

\section{Directional Bias}
While the Eddington inversion method provided a useful framework for exploring the connection between idealized density profiles $\rho(r) \propto r^{-k}$ and their corresponding velocity distributions, it is less naturally suited for problems where the boundary distribution is known explicitly at large radii. For the present analysis, where we aim to propagate the interstellar Maxwellian distribution inward from $r_{\rm out} \sim 10^5$ AU to the detection region near $r_{\max}=4$ AU, a Liouville mapping approach is more appropriate. Liouville’s theorem guarantees that the phase space density is conserved along collisionless trajectories and this allows us to start from the physically meaningful boundary condition $f(v,r_{\rm out})=f_{\rm ISM}(v)$ and map each velocity vector inward using the constants of motion, which in this case are orbital energy $\mathcal{E}$ and angular momentum $J$ \cite{lo1miller1988applications,lo2lindblad1959galactic}. This not only makes the boundary condition explicit, but also directly incorporates the dynamics of unbound, hyperbolic orbits characteristic of ISOs, together with gravitational focusing effects. In contrast, the Eddington inversion presumes an isotropic, steady-state bound population, making it best suited for illustrative explorations of how different slopes $k$ bias the velocity distribution. However, for the boundary-driven ISO problem, Liouville mapping provides the more self-consistent and physically motivated route.
\\
\\
We formulate the phase space analysis in terms of the constants of motion for the Kepler problem which are the specific orbital energy $\mathcal{E}$ and the specific angular momentum $J$. This change of coordinates greatly improves numerical stability because the relevant flux of low angular momentum trajectories can be represented explicitly. The starting point is the distribution function at infinity, which we model as a Maxwellian shifted by the solar motion and for simplicity we adopt an isotropic Maxwellian with one-dimensional velocity dispersion $\sigma$, so that the velocity distribution at infinity is 
\begin{equation*}
    f_\infty(v) \propto v^2 \exp\left(-\frac{v^2}{2\sigma^2}\right).
\end{equation*}
Transforming variables to the specific orbital energy 
\begin{equation}
    \mathcal{E} = \tfrac{1}{2} v_\infty^2,
\end{equation}
we obtain an effective one dimensional energy density 
\begin{equation}
    \phi(\mathcal{E}) \propto \sqrt{\mathcal{E}}  \exp\!\left(-\frac{\mathcal{E}}{\sigma^2}\right).
\end{equation}
The second conserved quantity is the specific angular momentum $J = r v_\perp$. At the outer boundary of the system, $r_{\rm out}$, the maximum allowed angular momentum for a given $\mathcal{E}$ is $J_{\rm max} \simeq r_{\rm out}\, v_\infty$. Closer to the Sun, at a given radius $r$, the kinematic constraint from $v_r^2 \geq 0$ imposes 
\begin{equation}
    J \leq J_{\rm cap}(\mathcal{E},r) = r \sqrt{2(\mathcal{E} + GM_\odot/r)}.
\end{equation}
Thus, the allowed domain in $(\mathcal{E},J)$ space is compact and bounded by $J_{\rm max}$ and $J_{\rm cap}$.
\\
\\
To represent the anisotropy required to enhance low angular momentum trajectories we introduced a weighting function $g_R(J)$ which suppresses high-$J$ orbits relative to low-$J$ ones. The family of functions we employed has the form
\begin{equation}
    g_R(J) \propto \left( 1 - \frac{J}{J_0(R)} \right)^{\alpha(R,a)} \quad \text{for } J \leq J_0(R),
\end{equation}
and $g_R(J)=0$ otherwise and here $J_0(R)$ sets the cutoff scale and is parameterized as
\begin{equation}
    J_0(R) = J_{0,{\rm base}} \left( \frac{R}{R_0} \right)^{-s},
\end{equation}
while the exponent is taken as $\alpha(R,a) = a \, \left( \tfrac{R}{R_0} - 1 \right)$, where $a$ is a tunable anisotropy amplitude calibrated to match the observational constraints. For $\alpha=0$, $g_R(J)$ is flat, while for $\alpha>0$, the function favors orbits with lower $J$.
\\
\\
Given this phase space distribution, the density of particles crossing a spherical shell at radius $r$ for a population of objects of physical radius $R$ is
\begin{equation}
    \rho(r;R,a) \propto \int_0^{\mathcal{E}_{\max}} d\mathcal{E}\, \phi(\mathcal{E})
\int_0^{J_{\rm max}(\mathcal{E},r)} dJ \; g_R(J)\, \frac{J}{v_r(r,\mathcal{E},J)} ,
\end{equation}
with radial velocity
\begin{equation}
    v_r(r,\mathcal{E},J) = \sqrt{2\big(\mathcal{E} + GM_\odot/r\big) - \frac{J^2}{r^2}}.
\end{equation}
To extract the effective slope $k(R)$ that measures the relative enhancement of encounter rates with object size, we compute $\rho(r;R,a)$ in a narrow logarithmic window around the detection radius $r_{\max}=4\,{\rm AU}$ and fit a local power law
\begin{equation}
    \rho(r;R,a) \propto r^{-k(R)}.
\end{equation}
This procedure gives an estimate of $k(R)$ by comparing $\rho$ at radii slightly above and below $r_{\max}$ and since only relative values matter, we normalize such that $k(R_0)=0$ at $R_0=0.6$ km, in accordance to \cite{loeb20253i}.
\\
\\
We then adjust the anisotropy amplitude $a$ to be in agreement with the required enhancement implied by the cumulative detection rates and once $a$ is fixed, the function $k(R)$ follows uniquely from the dynamics.
\begin{figure}[!h]
    \centering
    \includegraphics[width=1\linewidth]{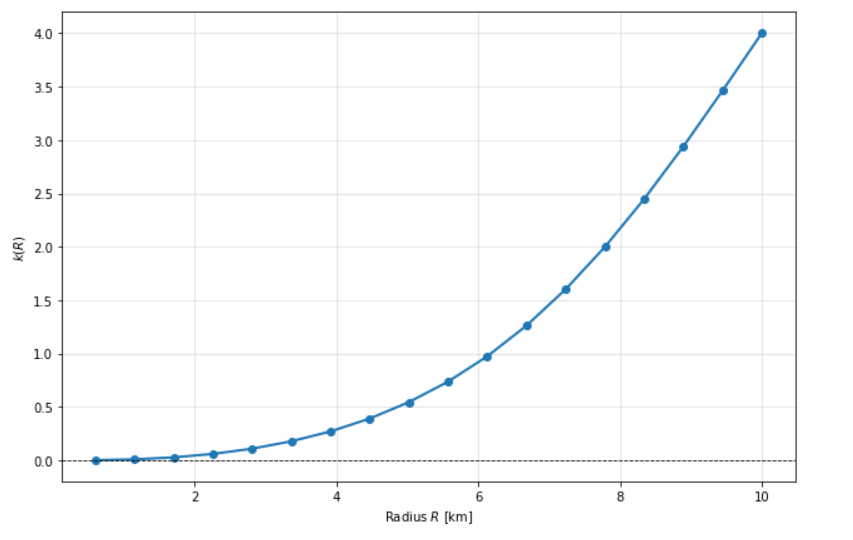}
    \caption{$k(R)$ as a function of ISO radius $R$, taking into account Liouville mapping formalism}
    \label{figkr}
\end{figure}
The resulting plot in Figure \ref{figkr} of $k(R)$ shows a smooth, monotonic rise from zero at $R=0.6$ km to $k(10)\approx 4$ at $R=10$ km. We can extend it to $R=23$km for the very large size mentioned in \cite{23lisse2025spherex}, but the trend remains the same. Physically, this curve quantifies how a bias toward low angular momentum trajectories, consistent with Liouville’s theorem, can reproduce the cumulative size dependence of the observed interstellar object detection rate. The fact that the curve is regular and monotonic indicates that the physics is internally consistent, and that the earlier numerical difficulties in direct velocity space integration were purely artifacts of coordinate choice. 
\\
\\
This curve makes an important point for the artificial origin discussion as it shows that within a physically consistent Liouville mapping in $\mathcal{E},J$ space, one can in fact generate the necessary enhancement in the encounter rate with increasing object size purely by favoring low‐angular momentum orbits. The bias needed to explain why large interstellar objects are detected at the observed rate can arise naturally from dynamical phase–space effects, without having to invoke any special mechanisms. The monotonic rise of k(R) to values $\sim 4$ is a signature that gravitational dynamics can in principle account for the observed scaling. But at the same time, the artificial scenario remains compelling as while the mapping shows that it is possible to match the observational size distribution slope by tuning the anisotropy in $J$, there is still no known astrophysical mechanism that would actually generate such a distribution of low angular momentum interstellar objects in the first place. This gap leaves room for speculation that the required anisotropy could be the result of artificial intervention, like from a directed release of objects with engineered orbits. 

\section{Conclusions}
In this work we have developed a dynamical framework to constrain the population of large ISOs using three complementary approaches. Encounter rate analysis demonstrated that detection of ISO with radii of tens of kilometers cannot be reconciled with a purely Maxwellian background, implying the need for significant flux enhancements. By applying the Eddington inversion method, we established how steep radial density profiles naturally correspond to phase-space distributions biased toward radial orbits, providing a clear dynamical mechanism for gravitational focusing near the Sun. Extending this discussion, we then employed a Liouville mapping in $(\mathcal{E},J)$ space to propagate the interstellar velocity distribution inward under explicit conservation laws, showing that the required size dependent enhancement of encounter rates can arise from anisotropies that favor low angular momentum trajectories.
\\
\\
These results suggest that the detection of ISOs with $R\gtrsim 10~{\rm km}$  is dynamically consistent with strong gravitational focusing, but only under specific velocity-space biases whose natural astrophysical origin remains unknown. This tension suggests two possible explanations, one being that new physical mechanisms like stellar perturbations or local structure in the Galactic environment produce the required velocity anisotropies. The other possibility is an orbital fine-tuning for ISOs of technological origin. Future ISO discoveries by the Vera C. Rubin Observatory \cite{vc1thomas2020vera,vc2blum2022snowmass2021,vc3sebag2020vera} will provide the crucial statistics needed to test these dynamical characteristics.

   \begin{acknowledgements}  OT was supported in part by the Vanderbilt Discovery Alliance Fellowship. AL was supported in part by the Black Hole Initiative, which is funded by GBMF anf JTF.
\end{acknowledgements}

\bibliography{references}
\bibliographystyle{aa}

\end{document}